\documentclass[12pt]{iopart}

\usepackage[dvips]{graphicx}
\usepackage{iopams}

\begin{document}

\title{Criticality of spreading dynamics in hierarchical cluster networks without inhibition}

\author{M Kaiser$^{1,2*}$, M G\"orner$^3$ and C C Hilgetag$^4$}
\address{$^1$ Newcastle University, School of Computing Science, Newcastle upon Tyne, NE1 7RU, United Kingdom}
\address{$^2$ Newcastle University, Institute of Neuroscience, Newcastle upon Tyne, NE2 4HH, United Kingdom}
\address{$^3$ UC Berkeley, Department of Mathematics, Berkeley, CA 94720, USA}
\address{$^4$ Jacobs University Bremen, School of Engineering and Science, Campus Ring 6, 28759 Bremen, Germany}
\ead{M.Kaiser@ncl.ac.uk}

\begin{abstract}
An essential requirement for the representation of functional patterns in complex neural networks, such as the mammalian cerebral cortex, is the existence of stable network activations within a limited critical range. In this range, the activity of neural populations in the network persists between the extremes of quickly dying out, or activating the whole network.
The nerve fiber network of the mammalian cerebral cortex possesses a modular organization extending across several levels of organization. Using a basic spreading model without inhibition, we investigated how functional activations of nodes propagate through such a hierarchically clustered network.  The simulations demonstrated that persistent and scalable activation could be produced in clustered networks, but not in random networks of the same size. Moreover, the parameter range yielding critical activations was substantially larger in hierarchical cluster networks than in small-world networks of the same size. These findings indicate that a hierarchical cluster architecture may provide the structural basis for the stable and diverse functional patterns observed in cortical networks.
\end{abstract}

\maketitle

\section{Introduction}
Natural systems operate within a critical functional range, sustaining diverse dynamical states \cite{Bak1987,Newman2005} on the basis of their intricate system architecture. For instance, in neural systems, such as the cerebral cortical networks of the mammalian brain, this critical range is indicated by the fact that initial activations result in various neuronal activity patterns that are neither dying out too quickly, nor spreading across the entire network. What are the essential structural and functional parameters that allow complex neural networks to maintain such a dynamic balance? In particular, which factors limit the spreading of persistent neural activity to the whole brain, thus preventing a pathological state resembling epilepsy?

Most current models of neural network dynamics focus on maintaining the right balance of network activation and rest through functional interactions among populations of inhibitory and excitatory nodes \cite{Beggs2003,Haider2006}. However, the topology of the networks may also make a significant contribution towards critical network dynamics, even in the absence of inhibitory nodes.

It is known from the anatomy of the brain that cortical architecture and connections are organized in a hierarchical and modular way, from cellular microcircuits in cortical columns \cite{Binzegger2004} at the lowest level, via cortical areas at the intermediate level, to clusters of highly connected brain regions at the global systems level \cite{Hilgetag2000b,Hilgetag2004,Sporns2004}. At each level clusters arise, with denser connectivity within than between the modules. This means that neurons within a column, area or area cluster are more frequently linked with each other than with neurons in the rest of the network.
We used a basic spreading model to explore the functional potential of different network topologies, particularly those possessing a clustered organization, for producing persistent yet contained node activations. The approach was inspired by classical 'netlet' studies of random neural networks \cite{Anninos1970}.  More recently, spreading analysis has also been applied to cortical networks \cite{Kotter2000} and to other complex networks with a non-random organization \cite{Pastor-Satorras2001,Dezso2002,Hufnagel2004}.

The present model operates without explicit inhibition, as we were specifically interested in the contributions of different network topologies to maintaining critical activations. This feature also reflects a structural attribute of cortical systems networks \cite{Latham2004} and other complex networks, for instance, social networks \cite{Pastor-Satorras2001}, that do not possess explicit inhibition, at least at some level of organization. Thus, while the current model is inspired by structural and functional properties of neural networks, it may also be relevant for understanding spreading phenomena in other complex systems, such as disease or virus spreading in social or computer networks \cite{Dezso2002,Eguiluz2002,Holme2002b}.

\section{Materials and Methods}
All simulations were implemented in Matlab 7 (Mathworks Inc, Natick, USA). To investigate the effect of clustered network organization on the spread of activity, simulations were run on three different types of networks: random networks, small-world networks with a fixed fraction $p\in [0,1]$ of random connections, and hierarchically clustered networks. All networks were undirected graphs with $N=1,000$ vertices and $E=12,000$ edges.  To create the (non-hierarchical) small-world networks, we used a procedure similar to the algorithm presented in \cite{Watts1998}; see \cite{sw}.

\begin{figure}
\includegraphics[width=2in]{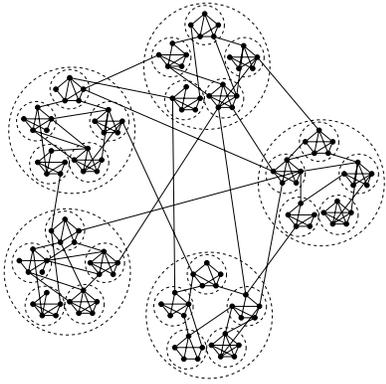}
\caption{Schematic view of a hierarchical cluster network with five clusters containing five sub-clusters each.\label{fig:schematic}}
\end{figure}

To create the hierarchical cluster network, the 1,000 vertices were parcellated into 10 disjoint sets ('clusters'), each consisting of 100 vertices.  Then each cluster was divided into 10 'sub-clusters', containing 10 vertices each. The network was wired randomly, such that 4,000 edges (one third of the total 12,000 connections) connected vertices within the same sub-clusters, further 4,000 edges connected vertices within the same clusters, and the last 4,000 were randomly distributed over all nodes of the network (Fig. \ref{fig:schematic}).

A basic spreading model \cite{Newman2005} was modified to simulate the spread of activity through the network. Note that the labels of activated/inactivated neural nodes, as used here, could be easily replaced by labels for infected/susceptible nodes in an analogous model of disease spreading.

The simulation operated in discrete time steps, with nodes being in one of two states: active or inactive. To advance from one time step to the next:
\begin{itemize}
\item An inactive node became activated in the next time step, if it was connected to at least $k$ (default $k=6$) active nodes.
\item An active node became inactivated with probability $\nu$ (default $\nu=0.3$) in the next time step, or otherwise stayed active.
\end{itemize}
For initialization, $i$ ($i\leq i_0$) nodes among the nodes $1$ to $i_0$ were selected randomly and activated in the first time step.  The networks nodes were numbered consecutively. Hence, by setting $i_0$ to, for example, 10, 20 or 100, only nodes in the first sub-cluster, the first two sub-clusters, or the first cluster, respectively, were activated during initialization.  Thus, $i$ determined the number of initially active nodes while  $i_0$ controlled the localization of initial activations, with smaller values resulting in more localized initial activity.

\begin{figure*}
\includegraphics[width=\textwidth]{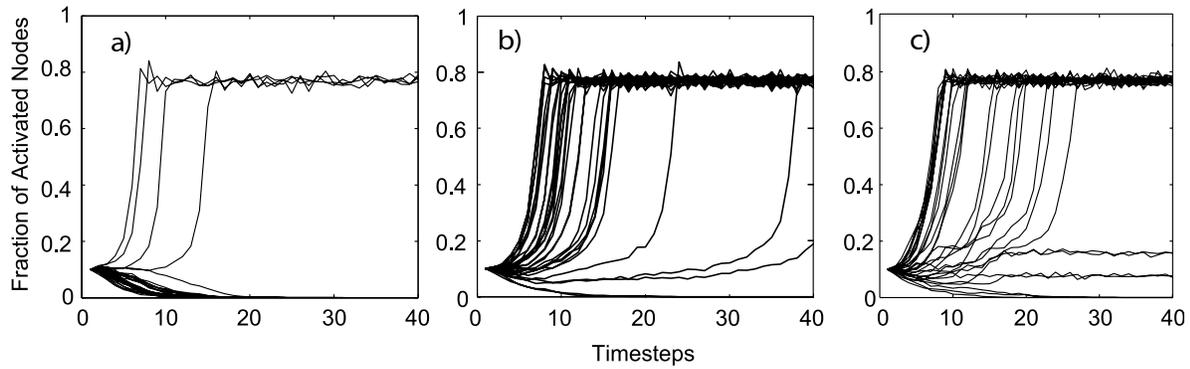}
\caption{Spread of initially unlocalized activity in (a) random, (b) small-world and (c) hierarchical cluster networks ($i=100, i_0=1,000$), based on 30 simulations for each network.
\label{fig:main}}
\end{figure*}

For a systematic comparison of the criticality behavior of different network topologies, small-world networks were tested for different fractions of randomly wired links, $p$.  When plotting Fig. \ref{fig:main} for various values of $p$, the simulation yielded sustained intermediate activity (neither dying out nor spreading through the whole network) only if $p$ was between 0.4 and 0.6. Consequently we chose $p=0.5$ for all simulations. Note that for this $p$, the small-world and hierarchical network had similar Clustering Coefficients and Characteristic Path Lengths (see table \ref{table:topology}). This means that the hierarchical cluster network also possessed small-world characteristics \cite{Watts1998}.  Given the high probability of rewiring in the default small-world network, one may wonder if this network still possesses strong {\it local} clustering comparable to the hierarchical cluster network. To test this aspect, we measured the edge density within each consecutive set of 10 network nodes in the small-world network, that is for 1,000 sets in total. The average local edge density in the small-world network was 0.51$\pm$0.08, and thus comparable to the average local edge density of 0.60$\pm$0.15 measured in the same way in the hierarchical cluster network.

\section{Results}
Across different simulation conditions, the hierarchical cluster network showed a larger variety of behaviors than the random or small-world networks, and it produced persistent yet contained network activity for a wider range of initial activation conditions.  Moreover, in contrast to random networks, the activation behavior of the hierarchical cluster network was also influenced by the localization of the initial activity.

\begin{figure}
\includegraphics[width=\textwidth]{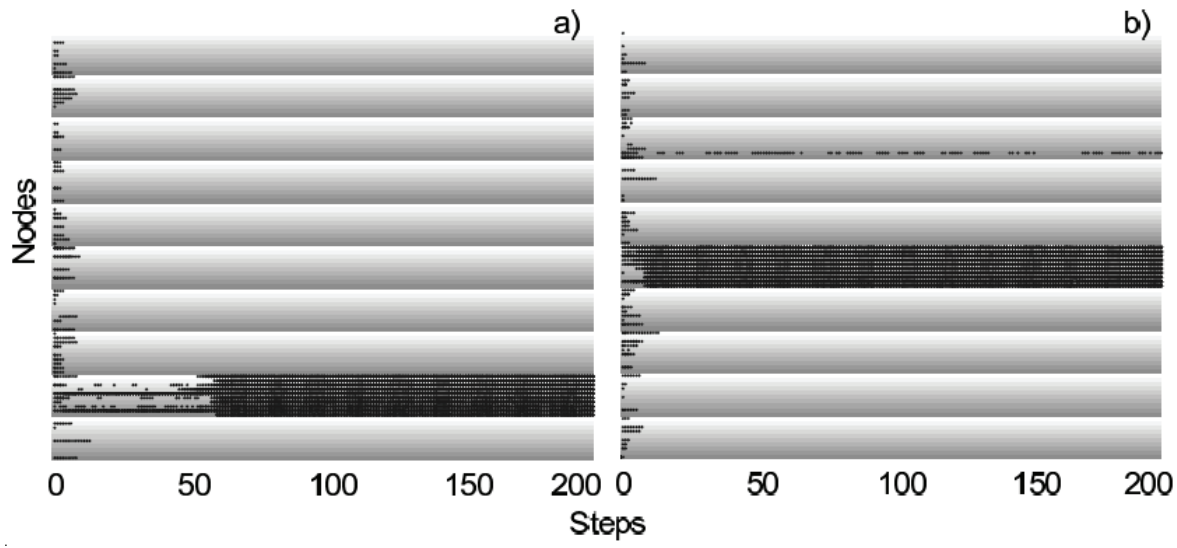}
\caption{Examples for different sustained activity patterns in hierarchical cluster networks ($i=90, i_0=1,000$). Graded gray background shading indicates the 10 subclusters within each of the 10 clusters. Black dots represent nodes active at the respective time step. (a) one cluster showing sustained activity. (b) one cluster remaining active with frequent co-activation of one external subcluster. \label{fig:evolution}}
\end{figure}

Examples exhibiting the behaviors of the different networks are given in Figure \ref{fig:main}.  The figure shows the result of 30 simulations in all three networks when 10\% of the nodes were randomly selected for initial activation. In the random network, activity either quickly spread through the network, or, more frequently, died out. In the small-world network, spread of activity was slower than in the random network, but eventually also resulted in all-or-nothing activation of the network. Compared to the random network, activation of the whole network was more frequent.  In contrast to the previous two types of networks, the hierarchical cluster network showed a large variety of delays for the spread of activity and also produced cases in which spreading was limited to one or two clusters with sustained activity.  Such persistent activation could be sustained with different patterns and varying extent of involved nodes (Fig.~\ref{fig:evolution}).

\subsection{Critical Range of Nonlocalized Activation}
To quantify the critical range of network function, we considered the range of parameters for which initial activations could result in intermediate levels of network activity and not just all-or-none network activation.

\begin{figure}
\includegraphics[width=3.41in]{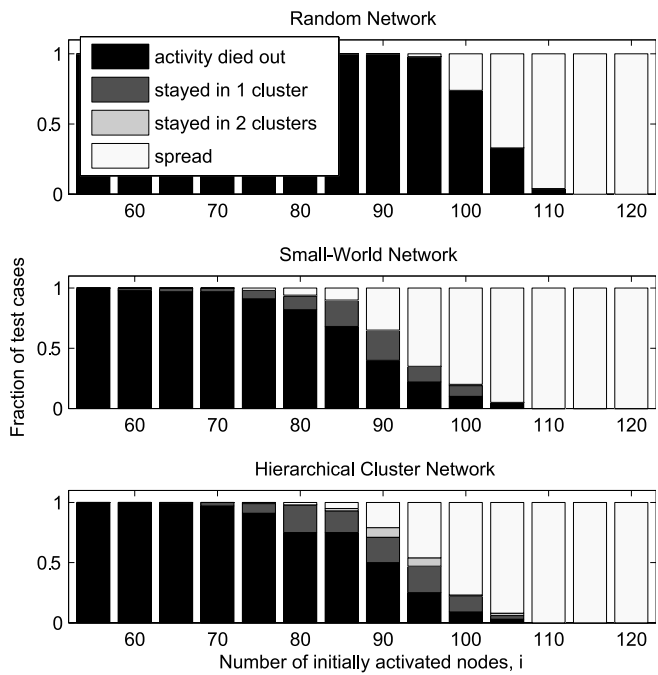}
\caption{Critical range of unlocalized node activations: Fraction of test cases resulting in different activation scenarios after 80 time steps versus number ($i$) of  initially activated, nonlocalized ($i_0=1,000$) nodes.\label{fig:criticalrange}}
\end{figure}

The critical range for each network was estimated by simulating the spreading of initially unlocalized activity ($i_0=1,000$) over 80 time steps for varying values of $i=40, 50, \dots 120$, while counting how many of the 100 simulations for a specific network and values of $i$ ended in one of the following scenarios: the number of activated nodes was zero (activity died out), between zero and 100 (activity stayed in one cluster), between 100 and 200 (activity stayed in two clusters), or larger than 200 (activity spread over the whole network); see Fig. \ref{fig:criticalrange}.
The critical range for initially unlocalized activity extended from around $i=65$ to $i=105$ in the case of the small-world and clustered hierarchical networks, while the random network only showed all-or-none activation.

\subsection{Influence of Localization}
\begin{figure}
\includegraphics[width=3.41in]{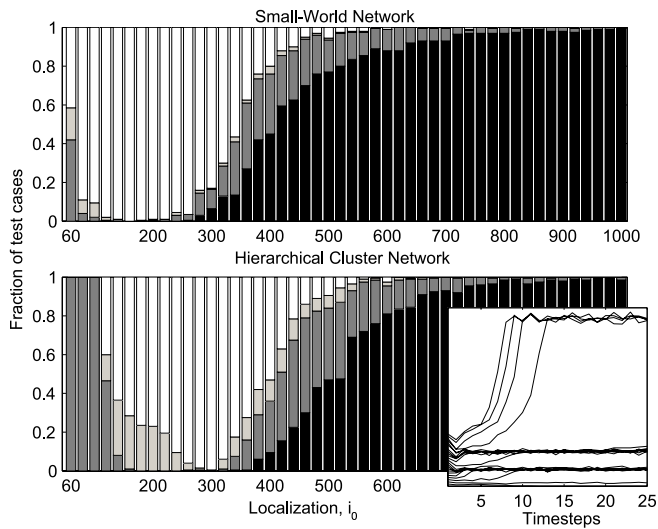}
\caption{Critical range of initially localized node activations: Fraction of test cases ending in different activation scenarios  after 80 time steps (see Fig. \ref{fig:criticalrange} for legend; gray levels for small-world networks represent activation in the neighborhood of the initially activated nodes) versus localization ($i_0=60, 80, \dots, 1,000$) of initially activated nodes.  Number of initially activated nodes was fixed at $i=60$. Data not shown for random networks, in which localization had no influence on the spread of activity and activity always died out for $i=60$.
  Sustained activity mostly remaining in clusters (inset): for each value of $i=0, 10, \dots, 220$, one simulation of localized activity ($i_0=i$) for 25 time steps is plotted for the hierarchical cluster network.
  \label{fig:localization}}
\end{figure}

Initially activated nodes may be localized in only a few clusters or distributed over the whole network. To investigate the impact of initial localization on the spread of activity, the simulations from the previous section were repeated with fixed $i=60$ and for values of $i_0$ ranging from 60 to 1,000 in steps of 20 (Fig. \ref{fig:localization}).

For the random network, activity always died out for $i=60$. Thus, localization had no impact on spreading in this type of network. The spread of activity in small-world and hierarchical cluster networks, however, depended on where the initial activation was localized. When initially active nodes were widely dispersed, activity in the network would eventually die out. More focused initial activation, on the other hand, could activate the whole network, or lead to persistent activation of network subregions.  This scaled partial activation was particularly pronounced in the hierarchical cluster network (example given in inset of Fig. \ref{fig:localization}).

The simulations also indicated a localization effect for the small-world network. The  plots in Fig. \ref{fig:localization} for the small-world and clustered hierarchical networks appear similar, but shifted, with a larger range of intermediate activations in the hierarchical cluster network. Notice, however, that the appearance of intermediate activation in small-world networks may have been partly due to the very slow spreading of activity in this type of network. Thus, many settings eventually leading to complete network activation had not yet activated the whole network at the end of the observation period and were therefore classified as partial activation.

\subsection{Delay of Activation Spreading}
We also investigated the time needed to activate 50\% of the nodes in a network, considering only those test cases in which activity did not die out.

In the random network, if activity spread at all, it did so quickly, typically in less than 10 time steps. Even if the initial activity was in the borderline range for all-or-none network activation, not more than 15 time steps were required in any of the cases. This was in contrast to the small-world and hierarchically clustered networks, for which a wide range of delay times was observed.
For the small-world network, delayed spreading could be produced by selecting the initial activity in the critical range, as shown in Fig. \ref{fig:criticalrange}. However it was also important that the initial activity was strictly localized ($i_0=i$). Setting $i_0=i=90$ typically resulted in about 40 time steps for spreading, whereas for $i_0=190, i=90$, spreading in the small-world network appeared similar to that in the random network.

By contrast, for the hierarchically clustered network, spreading to the global level did not arise when the initial activation was too strictly localized (Fig. \ref{fig:localization}). A maximum delay for spreading was achieved by localizing the initial activity within two or three clusters (e.g., delay around 40 steps for $i_0=200, i=90$).

\subsection{Parameter Range for Persistent Contained Activity}
\begin{figure*}
\includegraphics[width=\textwidth]{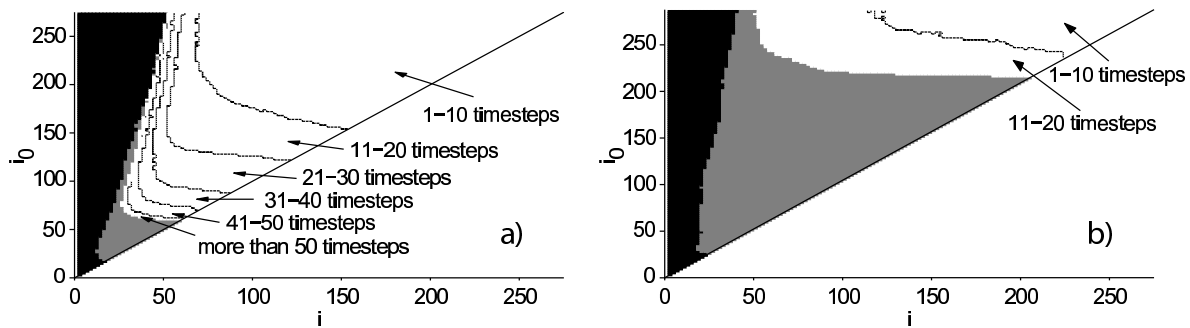}
\caption{Parameter space exploration of the critical range for all combinations of initial activation parameter $i$ and localization parameter $i_0$, based on 1,000 test cases. Simulation outcomes are indicated by graylevel (black: activity died out; gray: limited spreading; white: complete spreading). (a) small-world network. (b) hierarchical cluster network. \label{fig:phasespace}}
\end{figure*}

We systematically explored the network activation behaviors resulting from different settings of the initial node activation and localization parameters. Since initial activation is typically small in neural systems, we limited the maximum number of initially active nodes to 250, that is, one quarter of all network nodes. Persistent contained activity in hierarchical networks was robust for a wide range of initial localization and activation parameters (indicated by gray parameter domain in Fig. \ref{fig:phasespace}). For small-world networks, however, parameters needed to be finely tuned in order to yield sustained activity. Thus, hierarchical networks showed sustained activity for a wider range of initial conditions.

\begin{figure*}
\includegraphics[width=0.5\textwidth]{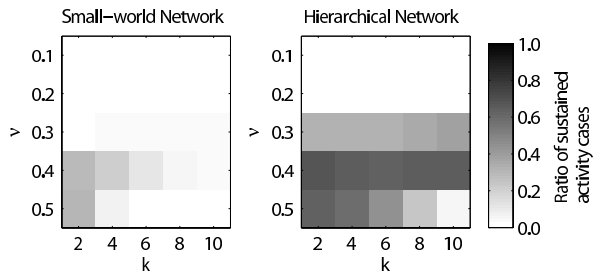}
\caption{Ratio of sustained activity cases depending on the spreading parameters $k$ (activation threshold) and $\nu$ (deactivation probability) for (a) small-world and (b) hierarchical cluster networks. \label{fig:parameterspace}}
\end{figure*}

We also explored if these results were robust for variations in the spreading parameters $k$ and $\nu$. For these control studies, we used a Monte Carlo approach that, for each pair of $k$ and $\nu$, generated 20 small-world and 20 hierarchical networks. For each network, the dynamics for 1,000 randomly chosen parameters  $i$ and $i_0$ were tested (Fig. \ref{fig:parameterspace}). A trial was considered to show sustained activity if $\geq$50\% of the nodes were activated at the end of the simulation. For each pair of spreading parameters $k$ and $\nu$, the average proportion of cases for which sustained activity occurred, which is related to the ratio of the gray space in Fig.~\ref{fig:phasespace}, was larger for the hierarchical cluster networks than for small-world networks. The maximum ratio for small-world networks was 30\%, whereas it was 67\% for hierarchical cluster networks.

\begin{figure*}
\includegraphics[width=0.5\textwidth]{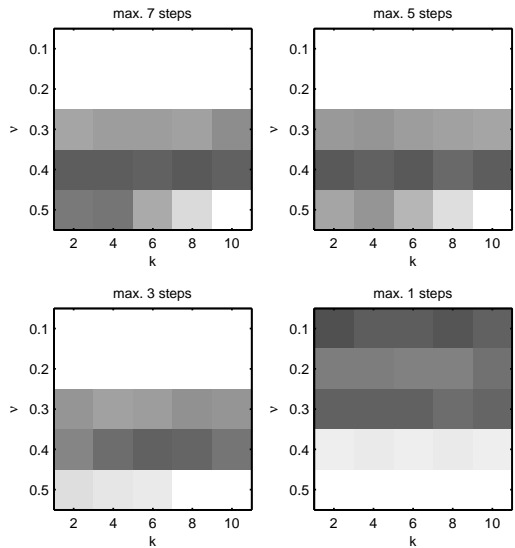}
\caption{Influence of exhaustion on sustained spreading for hierarchical cluster networks. As the maximum number of steps is reduced from seven down to one, sustained activity only occurs for lower values of $\nu$ whereas the total parameter range of sustained activity remains the same. Graylevels indicate the proportion of cases with sustained activity (cf. colorbar in Fig.~\ref{fig:parameterspace}). \label{fig:exhaustion}}
\end{figure*}

Finally, spreading of activity in the hierarchical cluster network also depended on the relative distribution of connections over sub-clusters, clusters and the whole network. To explore this effect, 1,000 trials were run for each of the connection distribution (Appendix, table \ref{tab:ChangeOfConnections}), with random values of $i$ and $i_0$, and $k=6$ and $\nu=0.3$. We measured the proportion of trials showing sustained activity (this approach was analogous to the one used for Figs. \ref{fig:parameterspace} and \ref{fig:exhaustion}).  Reducing the number of connections within clusters or sub-clusters blurred the boundaries of local network modules and reduced the proportion of cases with sustained activity, but the proportion was still larger than that for small-world networks. On the other hand, increasing the density of connections within sub-clusters and clusters led to a higher proportion of cases with sustained activity.

For the default arrangement of the hierarchical network (with 4,000 connections distributed at each of the three levels of organization), an activation of more than two clusters always resulted in spreading of activity through the whole network, due to the relatively high density of connections between clusters. Conversely, a reduction in the density of inter-cluster connections allowed persistent contained activity in more than two clusters (data not shown).  In these cases, the limited number of inter-cluster connections formed a bottleneck for activation of the remaining clusters.

In summary, the prevalence of persistent yet contained activity in hierarchical cluster networks was robust over a large range of model parameters and initial conditions, whereas in small-world networks, sustained activity occurred only for a relatively small parameter range.

\subsection{Exhaustion}
Energy resources for sustaining neural network activations may be limited. For instance, exhaustion occurs during epileptic seizures, reducing the duration of large-scale cortical activation. Similar phenomena arise in networks where sustaining an active/infected state or preserving the existence of nodes relies on limited resources.  Therefore, we also tested the effect of restricting the number of time steps that nodes could be consecutively active.

While in the previous analyses nodes could be active for an unlimited time, we here adjusted the number of simulation steps for which a node could be active without interruption, with upper limits from 7 to 1 steps (Fig.~\ref{fig:exhaustion}).  The diagrams show that persistent yet contained network activation could still occur in the hierarchical cluster network, despite different degrees of limiting node exhaustion.  As the number of steps was reduced, a lower value of the deactivation parameter $\nu$ was sufficient to keep activation in balance and allow sustained activity. Sustained activity was, however, largely independent of the exhaustion threshold parameter. The range of parameters for which sustained activity occurred remained similar to that in the previous analyses, with no clear correlation to the number of steps (average ratio of sustained activity cases over all pairs of the spreading parameters was 0.272$\pm$0.068).

\section{Discussion}
Our simulations demonstrated a strong impact of network topology on spreading behavior. Clustered networks were more easily activated than random networks of the same size.  This was due to the higher density of connections within the clusters which facilitated local activation.  At the same time, the sparser connectivity between clusters prevented the spreading of activity across the whole network.  Thus, in contrast to random networks, clustered networks possessed an expanded critical functional range for which initial activations resulted in persistent but non-global network activity.

In addition, there were also differences in the dynamical behavior of the two clustered networks types.  Compared to the small-world network, the hierarchical cluster network possessed a substantially larger critical parameter space, defined by the extent of localizations and initial network activations that could result in limited yet sustained network activity patterns (Fig. \ref{fig:phasespace}).

Simulations with similar results as described here were also performed for networks in which all nodes  were represented by integrate-and-fire (IF) neurons \cite{Koch1999}. These simulations also showed easier activation of the hierarchical cluster network compared to the random network, as well as the existence of intermediate states of activation in the hierarchical cluster, but not the random network \cite{Jucikas2006}. Thus, our results do not appear to depend on the specific neuronal model, but are generally based on the topology of the studied networks.  This conclusion is also supported by the systematic variation of model parameters presented here, which did not affect the principal model behavior.

The present hierarchical cluster model, which reflects the distributed multi-level modularity found in biological neural networks, is different from 'centralistic' hierarchical modular networks (with most nodes linked to network hubs) that have been described previously \cite{Ravasz2002}. While developmental algorithms have been suggested for the latter type of network, there are currently no algorithms for producing the hierarchical cluster networks presented here. However, single-level clustered network architectures can be produced by models for developmental spatial growth \cite{Kaiser2004b} or dynamic  self-organization of neural networks \cite{Izhikevich2004b}; and such models may serve as a starting point for exploring the biological mechanisms for developing multi-level clustered neural architectures.

In conclusion, the present results provide a proof of concept for three points. First, persistent but contained network activation can occur in the absence of inhibitory nodes. This might explain why cortical activity does not normally spread to the whole brain, even though top-level links between cortical areas are exclusively formed by excitatory fibers \cite{Latham2004}. While the involvement of inhibitory neurons and other dynamic control mechanisms may further extend the critical range, the present results indicate that the hierarchical cluster architecture of complex neural networks, such as the mammalian cortex, may provide the principal structural basis for their stable and scalable functional patterns. Second, in hierarchical clustered networks, activity can be sustained without the need for random input or noise as an external driving force. Third, multiple clusters in a network influence activity spreading in two ways: bottleneck connections between clusters limit global spreading whereas a higher connection density within clusters sustains recurrent local activity.

These findings may have practical implications and may guide future research.  For instance, it might be worthwhile to test whether epileptic patients show a higher degree of connectivity between cortical network clusters or other changes in structural connectivity which would facilitate spreading. Such changes might be reflected in certain aspects of functional connectivity \cite{Salvador2005} or be demonstrated more directly by diffusion tensor imaging. It would also be interesting to evaluate the impact of inhibition added to the lowest level of the network organization, such as provided by inhibitory interneurons in the cerebral cortex. However, by abstracting from specific characteristics of biological neural networks, the present study may also help to elucidate general features of spreading behavior in other hierarchically organized networks without limiting it to neural systems.

\ack
We thank Tadas Jucikas for performing the simulations with IF networks and Bernhard Kramer for helpful comments on this manuscript. M.K. and M.G. acknowledge financial support from the German National Merit Foundation. M.K. was supported by EPSRC (EP/E002331/1) and Royal Society (2006/R2).

\appendix
\section*{\label{apx}Appendix}
{\bf Spreading of activity depending on the relative density of connections within sub-clusters, clusters and at the network level in the hierarchical cluster network} \\
Simulating activity spreading for $k=6$ and $\nu=0.3$, sustained activity occurred in 18.4\% of the cases for 1,000 sub-cluster and 7,000 global connections and in 7.9\% of the cases for 1,000 cluster and 7,000 global connections (Table \ref{tab:ChangeOfConnections}). These proportions are lower than for the original network arrangement with 4,000 cluster, sub-cluster, and global connections, resulting in 43.6\% of cases with sustained activity, but still higher than the respective proportion of 1.96\% for small-world networks.

On the other hand, the number of cases with sustained activity was increased, if relatively more connections existed within clusters or sub-clusters. For example, increasing edge density within sub-clusters from 90\% to 98\% led to sustained activity in 53.7\% of the cases. Increasing edge density within clusters from 17\% to 21\% led to sustained activity in 85.1\% of the cases.

\newpage

\begin{table*}[h]
\caption{\label{table:topology}Topological properties of the networks}
\begin{indented}
\item[]\begin{tabular}{@{}cccc}
\br
& Random  & Small-world & Clustered\\
Quantity & Network & Network & Network\\
\mr
Clustering Coefficient & 0.025 & 0.11 & 0.15\\
Characteristic Path Length & 2.5 & 2.6 & 2.6 \\
\br
\end{tabular}
\end{indented}
\end{table*}

\begin{table*} 
\caption{\label{tab:ChangeOfConnections}Proportion of cases with sustained activity for different ratios of connections within clusters or sub-clusters. Note that the total number of connections within the network remained unchanged.}
	\begin{indented}
		\item[]\begin{tabular}{@{}lllll}
		\br
	& Network  & Connections & Connections  & Proportion of\\
&connections & within      & within       & cases with\\
&            & clusters    & sub-clusters & sustained activity\\
\mr
Original calculations	& 4,000	& 4,000	& 4,000	& 0.436\\
Increased cluster/ \\ network ratio &	3,600	&4,400	&4,000	&0.546\\
"	&2,000	&6,000	&4,000	&0.851\\
Reduced cluster/ \\ network ratio	&5,000	&3,000	&4,000	&0.183\\
"	&6,000	&2,000	&4,000	&0.159\\
"	&7,000	&1,000	&4,000	&0.079\\
Increased sub-cluster/ \\ network ratio	&3,600	&4,000	&4,400	&0.537\\
Reduced sub-cluster/ \\ network ratio	&5,000	&4,000	&3,000	&0.203\\
"	&6,000	&4,000	&2,000	&0.237\\
"	&7,000	&4,000	&1,000	&0.184\\
\br
		\end{tabular}
	\end{indented}
\end{table*}


\begin{thebibliography}{20}
\bibitem{sw}
{The algorithm started with a regular ring lattice where the vertices were placed on a circle and each vertex was connected to its nearest neighbors, its second nearest neighbors, \dots and its $d$-th nearest neighbor, where $d$ was obtained by rounding $E/N$ to the next integer. Then $(1-p)E$ edges were randomly picked from the regular ring lattice and $pE$ random edges were added to them to obtain a small-world network with $E$ edges.}
\end{thebibliography}
\end{document}